# Identification of constitutive parameters from full thermal and kinematic fields: application to hyperelasticity

S. Charlès[1], J.-B. Le Cam[1]

1. Univ Rennes, CNRS, IPR (Institut de Physique de Rennes) - UMR 6251, F-35000 Rennes, France

## ABSTRACT

In this paper, a new inverse identification method is developed from full kinematic and thermal field measurements. It consists in reconstructing the heat source from two approaches, a first one that requires the measurement of the temperature field and the value of the thermophysical parameters, and a second one based on the measurement of the kinematics field and a thermo-hyperelastic model that contains the parameters to be identified. The identification does not require any boundary conditions since it is carried out at the local scale. In the present work, the method is applied to the identification of hyperelastic parameters from a heterogeneous heat source field. Due to large deformation undergone by the rubber specimen tested, a motion compensation technique is developed to plot the kinematic and thermal fields at the same points before reconstructing the heat source.

## INTRODUCTION

Several methods have been recently developed for identifying parameters from field measurements. They are reviewed in [1]. In many of these approaches, the boundary conditions are necessary to solve the identification problem. The present study aims at developing a methodology for inverse identification using only local quantities. This means that constitutive parameters would be identified from a zone at the surface of the specimen, whatever the loading conditions applied to it. This implies that local quantities explicitly depend on the strain-stress relationship. In this work, we propose to identify the constitutive parameters by reconstructing the heat source field according to two different ways: a first one that requires the kinematic field and a given thermomechanical model that contains the parameters to be identified, and a second one that needs the temperature field and the thermophysical parameters. This inverse identification method has been applied to a hyperelastic material, which involves several difficulties. Indeed, hyperelasticity is generally used as a first approximation to predict the mechanical response of rubbery materials while several phenomena come into play in the deformation process. Numerous constitutive relations are available in the literature and reviewed in [2]. Due to the fact that hyperelastic models do not account for the above-mentioned phenomena, the values of the hyperelastic constitutive parameters depend on the strain state. This is the reason why constitutive parameters are classically identified from several homogeneous tests, namely uniaxial tensile (UT), pure shear (PS) and equibiaxial tensile (EQT). These three tests completely describe the domain of possible loading paths. A trade-off between the sets of values obtained with the different tests has therefore to be found to obtain parameters that can reasonably be considered as intrinsic to the mechanical behaviour of the material. Such identification approach exhibits several disadvantages, such as the necessity of making different geometry for the different tests, and the comparison between the constitutive parameters identified from the different loadings. An alternative approach has been proposed, based on the fact that the identification of constitutive parameters can be done from only one heterogeneous test, as soon as it induces at least the three tests mentioned above. In fact, a wide range of loading is also induced [3,4]. In the present study, such heterogeneous test is used to identify the hyperelastic constitutive parameters from a heat source approach.

# HEAT SOURCE APPROACH

In this approach, the heat source field is reconstructed from two different ways. A first one is based on the kinematic field and a given thermomechanical model that contains the parameters to be identified, and a second one is based on the temperature field and the thermophysical parameters. Considering that the constitutive state equations derive from the Helmoltz free energy function and that heat conduction follows the Fourier's law, the local diffusion equation writes:

$$\rho_0 C T - Div(\boldsymbol{K_0} Grad T) - R = S$$

where
- $\rho_0$ is the density in the reference configuration,
- $C$ is the heat capacity,
- $\boldsymbol{K_0}$ is the thermal conductivity tensor,
- $T$ is the absolute temperature,
- $R$ is the external heat source (from radiation for instance),
- $S$ is the heat source in the Lagrangian configuration.

IR thermography provides in-plane full temperature fields. Therefore, the specimen under study has to be thin and a two-dimensional version of the heat equation is required to reconstruct the heat source field. For that purpose, several assumptions are used. First, the heat conduction is considered as isotropic. Second, the temperature is considered to be homogeneous through the specimen thickness. Third, the external radiations $R$ are assumed to remains constants over time. These assumptions leads to the two-dimensional formulation of the heat diffusion equation:

$$\rho_0 C (\dot{\theta} + \frac{\theta}{\tau} - k_0 \Delta_{2D} \theta) = S$$

where
- $k_0$ is the coefficient of thermal conductivity,
- $\theta$ is the temperature variation $\theta = T - T_0$,
- $\tau$ is a time characterizing the heat exchanges along the Z-direction by convection with the air at the specimen's surface,
- $\Delta_{2D}$ is the Laplacian operator in the specimen plane in the Lagrangian configuration.

The prediction of the heat sources produced during the deformation process requires the choice of a free energy. Here, the material is assumed to behave as a hyperelastic material that is mechanically incompressible and isotropic. At low strain levels, typically inferior to 250%, the Neo-Hookean model can be chosen to predict the mechanical behaviour [5]. The free energy function is then given by the following strain energy density:

$$W(\boldsymbol{F}, T) = \frac{1}{2} NkT (I_1 - 3)$$

where
- $\boldsymbol{F}$ is the deformation gradient tensor,
- N is the number of network chains per unit volume
- $k$ is the Boltzmann's constant,
- $I_1$ is the first invariant of the left Cauchy-green deformation tensor $\boldsymbol{B}$.

While the material is supposed to be incompressible, the deformation gradient tensor for a biaxial loading is given by:

$$\boldsymbol{F} = \lambda e_1 \otimes e_1 + \lambda^B e_2 \otimes e_2 + \lambda^{-(B+1)} e_3 \otimes e_3$$

where:
- $e_1$, $e_2$ and $e_3$ are three orthonormal vectors of the 3D Euclidian space,
- $\lambda$ is the stretch ratio in the $e_1$ direction,
- the operator $\otimes$ between two vectors is such that $[a \otimes b]_{ij} = a_i b_j$,
- $B$ is the biaxiality ratio. It is equal to -0.5, 0 and 1 for uniaxial tension, pure shear and equibiaxial tension, respectively.

In the case where the material does not produce intrinsic dissipation and no other thermomecanical couplings come into play, the heat source is given by:

$$S = NkT(\lambda + B\lambda^{2B-1} - (B+1)\lambda^{-2B-3})\frac{d\lambda}{dt}$$

It should be noted that for temperature variations that do not exceed a few degrees, the quantity $NkT$ remains nearly equal to $NkT_0$.

*Experimental setup*

In the present study, the material chosen is an unfilled nitrile rubber. Its thermomechanical behaviour is driven by the coupling between strain and temperature only, i.e. it does not produce intrinsic dissipation and no other thermomechanical couplings come into play. Therefore, only the contribution of the thermo-elastic coupling is considered in the heat source description. The specimen geometry is shown in Fig. 1(a).

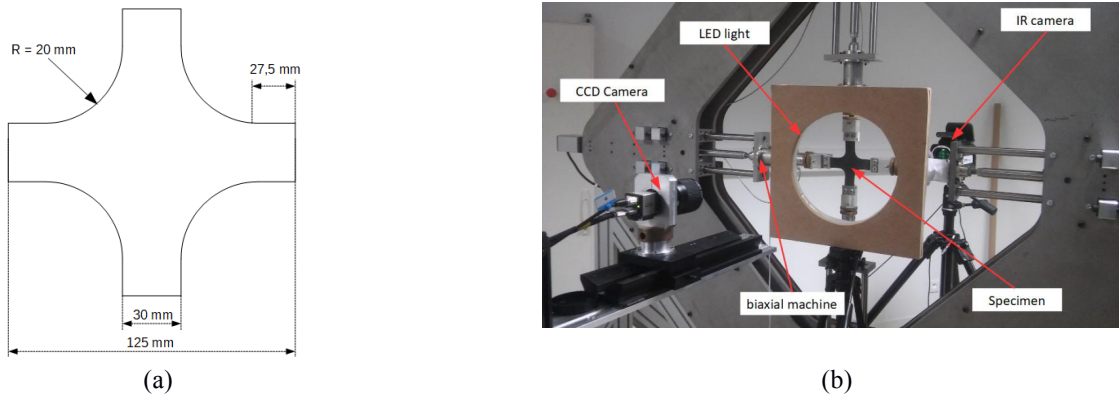

(a) (b)
Figure 1: (a) specimen geometry (b) overview of the experimental setup

Fig. 1(b) presents an overview of the experimental setup composed of an optical camera and an infrared one, on both side of the home-made biaxial testing machine. In the present study, an equibiaxial loading was applied, by controlling the four actuators. For each of them, the displacement and the loading rate was set to 70 mm and 150 mm/min, respectively. It should be noted that the two cameras are triggered for storing images at the same time, at the rate of 5 Hz. Displacement field at the specimen surface is determined by using the digital image correlation (DIC) technique. In order to improve the image contrast, a white paint is sprayed on the surface. The software used for the correlation process was SeptD [6]. The spatial resolution, defined as the smallest distance between two independent points, was equal to 10 pixels, which corresponds to 0.97 mm. The components of the deformation gradient tensor are determined at the centre of each square elements defined by the DIC grid. Within an element, the displacement is supposed to be a bilinear function of the Eulerian coordinates and can be obtain using the following equations:

$$\begin{cases} U_x(X_1, X_2) = a + bX_1 + cX_2 + dX_1X_2 \\ U_y(X_1, X_2) = e + fX_1 + gX_2 + hX_1X_2 \end{cases}$$

where $U$ is the displacement, are the Eulerian coordinates and *a, b, c, d, e, f, g* and *h* are constants that can be identified from the values at the four nodes. Finally, the components of deformation gradient tensor **F** are defined by $F_{ij} = \frac{x_i}{X_j}$, where $x$ are the Lagrangian coordinates. The three principal stretches ($\lambda_1 > \lambda_2 > \lambda_3$) are defined as the square roots of the eigenvalues of the left Cauchy-Green tensor **B** ($\boldsymbol{B} = \boldsymbol{FF}^T$). Since in-plane displacement is measured, only $\lambda_1$ and $\lambda_2$ are determined, $\lambda_3$ is deduced by assuming the material to be incompressible. The biaxiality coefficient *B*, defined as $\log(\lambda_2/\lambda_1)$, is then computed. The two invariants $I_1$ and $I_2$ of **B** are used to characterize the heterogeneity of the stretch states and the distribution in the maximum principal stretch value.

Temperature measurements were performed by using a FLIR infrared camera. The calibration of camera detectors was carried out with a black body using a one-point Non-Uniformity Correction (NUC) procedure at the acquisition frequency. The noise equivalent temperature difference (NETD) is equal to 20 mK for a range between 5 and 40°C. The spatial resolution of the thermal field was equal to 300 µm/pixel. In order to stabilize its internal temperature, the camera was switch on several hours before the test. Due to large deformation undergone by the material, materials points observed by the IR camera move from pixel to pixel in the IR images. The temperature variation at a given material point has therefore to be processed by compensating its movement and to track its temperature in the IR images [7-10]. This requires first describing the kinematic and thermal fields in the same coordinate system. For that purpose, a calibration test pattern is positioned in place of the specimen. It is a plate with two sets of four holes, one that forms a large rectangle and a second one that forms a small rectangle. This is illustrated in Figure 2. The set of holes used depends on the spatial resolution of the kinematic and thermal fields. Mark tracking algorithm is used to determine coordinates of these holes for both optical and IR camera. After that, a shared coordinate system is defined by choosing one of these marks as the reference. Thanks to this shared coordinate system, each point where the deformation gradient tensor is computed can be plotted in the infrared camera. As the two cameras do not have the same resolution, the temperature of each point is interpolated from the four nearby IR pixels.

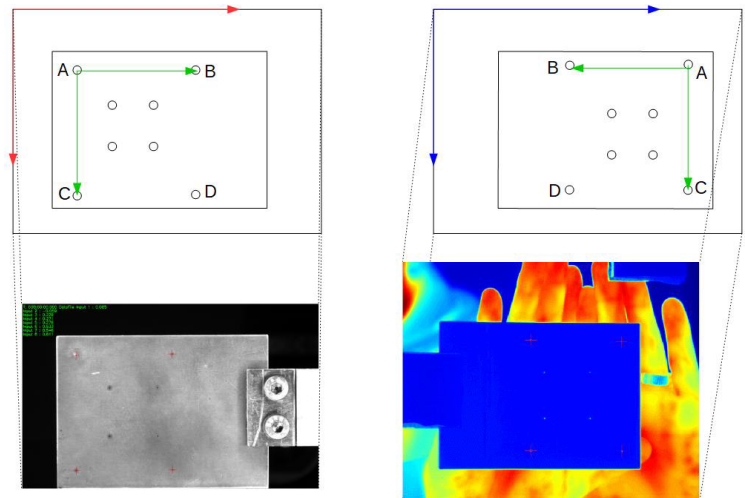

Figure 2: Scheme and image of the calibration pattern with the optical camera (on the left hand side) and with the IR camera (on the right hand side)

**RESULTS AND DISCUSSIONS**

The heterogeneous test presented in the previous section is performed. Figure 3 presents the displacement field along the x and y axis. Then, the deformation gradient tensor components are determined by using the methodology described in the previous section. The heterogeneity of the test can be evaluated by mapping the stretch states over the area of study. A colour scale is defined in such a way that ET, PS and UT states appear in blue, green and red colours, respectively. Figure 4 highlights the spatial distribution of the strain states: ET at the specimen center, UT in the branches and PS between these two states.

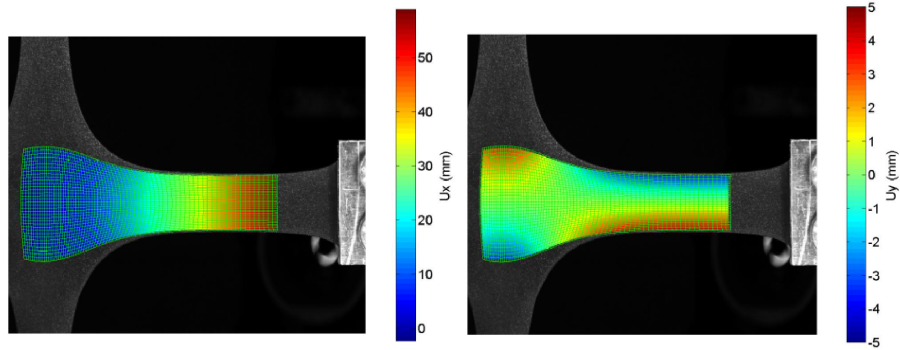

Figure 3: Displacement field along the x and y axis

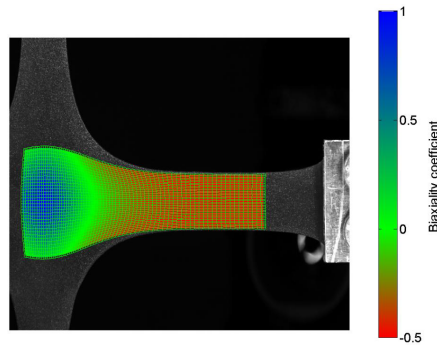

Figure 4: Biaxiality coefficient field

Motion compensation technique has been then applied to track the temperature of each point of the DIC grid. In order to reconstruct the heat source field from IR thermography measurements, the time constant $\tau$ has to be determined experimentally. The method used consists in heating the specimen and in measuring the temperature field during the return to thermal equilibrium, then in fitting the curve by an exponential function. Since heat source field can now be obtained both by the heat equation and the Neo-Hookean model, the constitutive parameter can be determined. Two different approaches can be considered, a global one considering an unique Neo-Hookean parameter for the whole specimen; a local one considering a different constitutive parameter for each ZOI studied. The results obtained will be precisely detailed and discussed in the presentation.

## CONCLUSION

In this paper, a new inverse identification is developed from the reconstruction of the heat source field from two approaches. The first one requires the measurement of the temperature field and the value of the thermophysical parameters. The second one requires the measurement of the kinematics field and the choice of a thermo-hyperelastic model that contains the parameters to be identified. Since it is a local approach, the boundary conditions are not needed. In the present work, the method is applied to large deformations of rubber, where the movement of the observed points has to be compensated. A motion compensation technique has been developed and validated. Finally, the identification is performed at the local scale and at the scale of the Region of Interest. A very good agreement in the force-displacement response measured during the test and predicted by a finite element simulation with the parameters identified illustrates the relevancy of this new inverse identification method from heat source field.

## ACKNOWLEDGEMENTS

The authors thank the National Center for Scientific Research (MRCT-CNRS and MI-CNRS), Rennes Metropole and Region Bretagne for financially supporting this work and the PCM Technologies S.A.S compagny for proiding the specimens. Authors also thank Dr Mathieu Miroir, M Vincent Burgaud and M Mickaël Lefur for having designed the biaxial tensile machine. Dr Eric Robin is acknowledged for providing the mark tracking algorithm.